\documentclass[epjCONF]{svjour}
\usepackage{graphicx}
\usepackage[varg]{txfonts} 
\usepackage[latin1]{inputenc}
\usepackage{color}
\session-title{%
19$^{\textnormal{\footnotesize th}}$ International %
IUPAP Conference on Few-Body Problems in Physics%
}
\newcommand{\vpr}{\vec{p}_{\rho}}
\newcommand{\vpl}{\vec{p}_{\lambda}}
\begin{document}
\title{%
Using highly excited baryons to catch the quark mass
}%
\author{%
T. Van Cauteren\inst{1}\fnmsep\thanks{\email{timvancauteren@gmail.com}} 
\and %
P. Bicudo\inst{2}
\and %
M. Cardoso\inst{2} 
\and %
Felipe J. Llanes-Estrada\inst{3} 
}
\institute{%
Dept. Physics and Astronomy, Ghent University, Proeftuinstraat 86,
B-9000 Gent, Belgium
\and
Dpto. F{\'i}sica, Instituto Superior T{\'e}cnico, Avda. Rovisco Pais, 
1096 Lisbon, Portugal
\and
Dpto. F{\'i}sica Te{\'o}rica I, Universidad Complutense, Avda. Complutense s/n,
28040 Madrid, Spain
}
\abstract{
Chiral symmetry in QCD can be simultaneously in Wigner and Goldstone modes, 
depending on the part of the spectrum examined.
The transition regime between both, exploiting for example the onset of
parity doubling in the high baryon spectrum, can be used
to probe the running quark mass in the mid-IR power-law regime. 
In passing we also argue that three-quark states naturally group into
same-flavor quartets, split into two parity doublets, all splittings decreasing
high in the
spectrum. We propose that a measurement of masses of high-partial wave
$\Delta^*$ resonances should be sufficient to unambiguously establish
the approximate degeneracy and see the quark mass running. 
We test these concepts with the first computation of the spectrum of high-J
excited baryons in a chiral-invariant quark model. }
\maketitle
%
%
%
\section{Introduction}\label{sec:intro}

In the last thirty years, Quantum Chromodynamics has turned out to give an
accurate description of many high-energy processes. In the last decennium,
a lot of effort has gone to solving the QCD equations on a discretized
spacetime lattice, using an impressive amount of computing power. This, together
with perturbation theory, has led
to very persuasive evidence that QCD indeed describes the strong interaction,
both at the high-energy end and at the static end. However, there still remains
the question if the theory describes the transition between the
perturbative and non-perturbative domain. 
A suitable quantity for checking this
is the quark mass, which should be a function of the quark's momentum due to
the interactions with the QCD medium. Asymptotic freedom tells us that at high
quark momentum $k$, the nonstrange quark mass should approach the current quark
mass of the order of $1-5$~MeV as deduced from chiral perturbation theory or
sum-rules, while hadron
phenomenology puts the value of the quark mass at low momenta to an
effective $\sim 300$~MeV. This means that the nonstrange quark mass changes by
two orders of magnitude when going from low momentum $k <<
\Lambda_{\textrm{QCD}}$ to high momentum $k >>
\Lambda_{\textrm{QCD}}$ (typically $\Lambda_{\textrm{QCD}} \sim 210$~MeV).

Obviously, a direct comparison between the QCD prediction of the running quark
mass and experimental observation is not possible since quarks are confined and
propagate only a distance of order a fermi, too small to detect directly. It is
a main goal of modern hadron physics to glimpse properties of the confined,
colored quarks from colorless hadron properties.
 We suggest an indirect way
of extracting the power-law behaviour of the quark mass function from the
experimentally obtainable masses of $\Delta$ resonances. Two main ideas are
applied
in order to do this: \emph{insensitivity to chiral symmetry breaking} leading to
the appearance of parity doublets in the hadron
spectrum~\cite{Glozman:1999tk,swanson,Wagenbrunn:2006cs} and \emph{linking the
mass splittings between parity partners with the running quark mass}. 

\begin{figure}[!h]
\centering
\includegraphics[width=0.75\columnwidth]{VanCauterenT-fig8.eps}
\includegraphics[width=0.75\columnwidth]{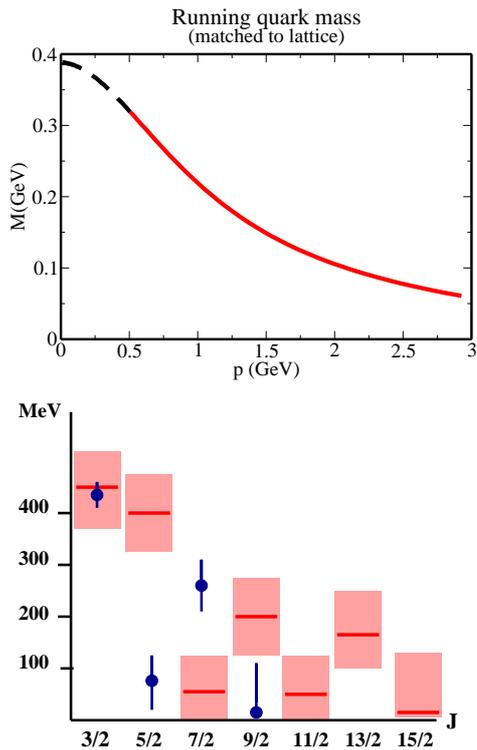}
\caption{
The running of the quark mass (top), falling with momentum as a power law, is
accessible from the hadron spectrum in the region $m(k) < \langle k \rangle$
(continuous red line). 
All one needs to do is fit a power law to the decreasing parity splittings for
excited baryon states (bottom, where we show a variational-Montecarlo model
calculation thereof as boxes and experimental data from Ref.~\cite{pdg} as
circles). Then the exponent of the quark mass power-law in Eq.~(\ref{key2}) is
obtained from Eq.~(\ref{key1}).
}
\label{fig:measureM}
\end{figure}

Our proposal can be concisely understood with the help of figure
\ref{fig:measureM}.

\section{Chiral quartets}\label{sec:chiral}

Chiral symmetry in the strong interactions is the Noether symmetry related to
the chiral transformation, which reads on the classical fermion fields
\begin{equation}
 \psi \to e^{i \alpha^a \tau^a \gamma_5} \psi \; , \label{eq:chiraltrf}
\end{equation}
with $\alpha^a$ the parameters describing the rotation in $SU(2)$-flavour space
with generators $\tau^a$. Due to the Dirac matrix $\gamma_5$, the chiral
transformation is a helicity-dependent rotation in flavour space.

One term in the QCD Lagrangian which is clearly not invariant to the chiral
transformation of Eq.~\ref{eq:chiraltrf}, is a quark mass term
\begin{equation}
 m_q \bar{q} q \; .
\end{equation}
This term breaks chiral symmetry explicitly, and the amount of breaking is
larger when the quark mass increases. As mentioned before, the small current
quark mass of $5$~MeV breaks chiral symmetry only slightly.  On top of the
explicit breaking of chiral symmetry, the strong interactions also show
spontaneous chiral symmetry breaking due to the appearance of the quark
condensate $<\bar{q} q>$ (a constituent
quark mass of $300$~MeV arises and breaks it to a much larger degree).

\begin{figure}[!h]
\centering
\includegraphics[width=0.75\columnwidth]{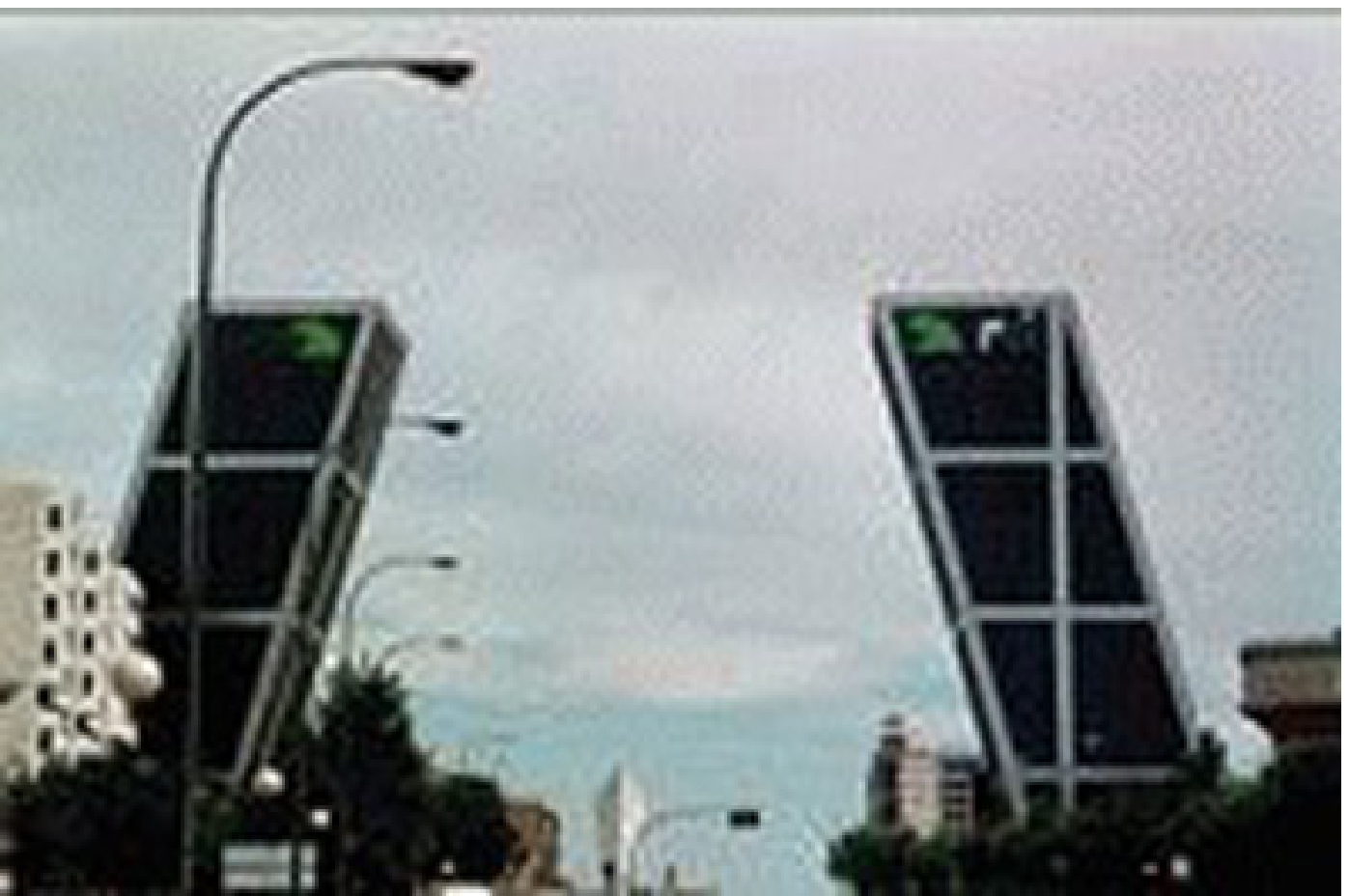}
\includegraphics[width=0.75\columnwidth]{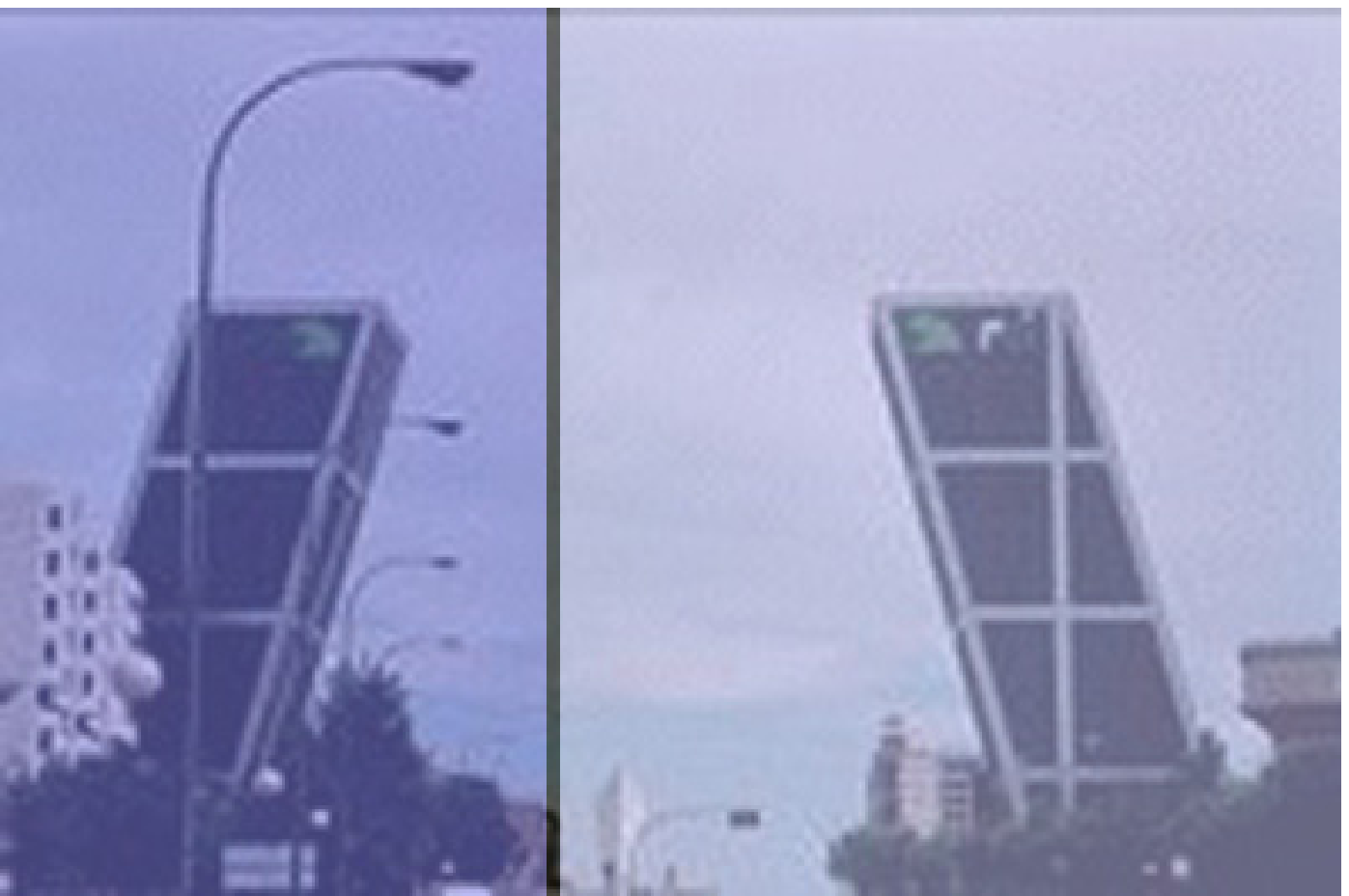}
\includegraphics[width=0.75\columnwidth]{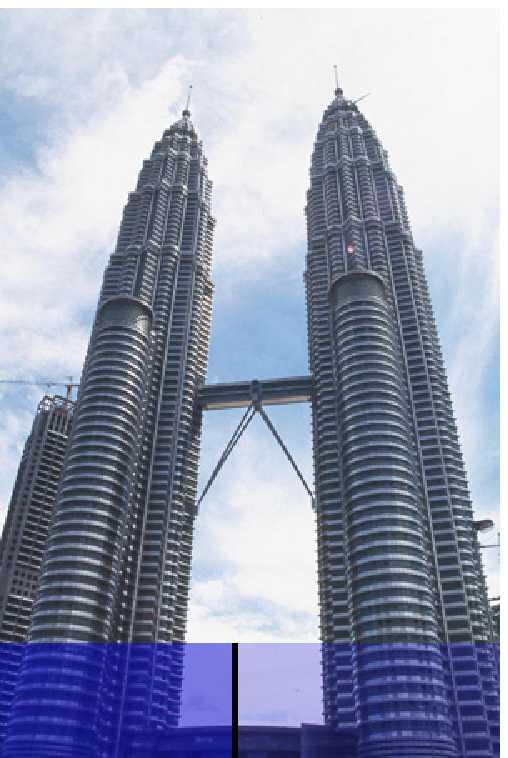}
\caption{
A classical analogy of three symmetry realizations in a quantum theory.
Top: the two buildings have equal mass as required by Reflection symmetry in
Wigner mode.
Middle: if inmersed in a fluid (non-trivial ground state) with a density
gradient from left to right, the two buildings, filled with it, have now
different mass (the symmetry is in Goldstone mode).
Bottom: even if the fluid breaks the Reflection symmetry, tall enough buildings
have close to equal masses (insensitivity to symmetry breaking). 
}
\label{fig:symmetrymodes}
\end{figure}

\begin{figure}[!h]
\centering
\vspace*{1cm}
\includegraphics[width=0.75\columnwidth]{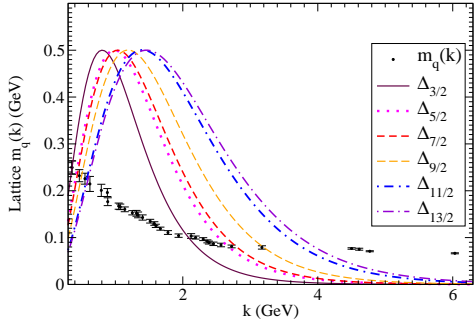}
\caption{Typical momentum distributions of increasingly excited $\Delta_{3/2},
\cdots, \Delta_{13/2}$ resonances overlap less and less with the dynamically
generated infrared quark mass. The momentum distributions are computed using
variational wave functions (not normalized for visibility) for a linear
potential with string tension $\sigma = 0.135$~GeV$^2$.}
\label{fig:masswithwf}
\end{figure}

This is the QCD explanation for the absence of parity doubling in the low
spectrum, as the symmetry is spontaneously broken. The community now  believes 
that the highly excited states however are insensitive to this spontaneous
breaking (see fig. 
\ref{fig:symmetrymodes} for a classical analogy). Indeed, since the quark mass
is large at low momenta and small at high momenta, one may
expect that the chiral symmetry breaking is less important in systems where the
average quark momentum $<k>$ is high, than in systems where $<k>$ is large.
This leads to the idea of insensitivity to chiral symmetry breaking high in the
hadron
spectrum
\cite{Glozman:1999tk,swanson,Wagenbrunn:2006cs,LeYaouanc:1984dr,Bicudo:1998bz} :
in highly excited meson or baryon states, the average
quark momentum can become larger than $\Lambda_{\textrm{QCD}}$ and the
explicit chiral-symmetry breaking quark mass term is small. This is illustrated
in Fig.~\ref{fig:masswithwf}, where typical quark momentum distributions are
shown for the lowest-lying $\Delta$-resonance for different spins, together
with the lattice QCD calculation of the running quark mass from
Ref.~\cite{bowman}.

\subsection{Series expansion in $m(k)/k$}\label{sec:expansion}

In order to assess the effect of a nonzero quark mass to a hadron mass, it is
important to note that the quark mass shows up both in the quark spinors and in
the QCD Hamiltonian. Since we want to find out what happens at small quark mass
and high momentum, a series expansion of the spinors and Hamiltonian in the
parameter $m(k)/k$ seems appropriate.

The spinors are expanded as
\begin{eqnarray}
 U_{k\lambda} = & &\frac{1}{2E(k)} \left[ \begin{array}{c} \sqrt{E(k) +
m(k)} \chi_\lambda \\ \sqrt{E(k) - m(k)} \vec{\sigma} \cdot \hat{k} \chi_\lambda
\end{array} \right] \\
\overrightarrow{k \to \infty} & & \frac{1}{\sqrt{2}} \left[ \begin{array}{c}
\chi_\lambda \\ \vec{\sigma} \cdot \hat{k} \chi_\lambda \end{array} \right] +
\frac{1}{2\sqrt{2}} \frac{m(k)}{k} \left[ \begin{array}{c} \chi_\lambda \\
-\vec{\sigma} \cdot \hat{k} \chi_\lambda \end{array} \right] \, , \nonumber
\end{eqnarray}
with $E(k) = \sqrt{k^2 + m(k)^2}$. We have kept the leading chirally invariant
term and the leading chiral-symmetry breaking term which is necessarily
of order $m(k)/k$. Note that the lower component of the first-order term has
the opposite sign of the lower component of the chiral invariant term.

When expanding the QCD Hamiltonian~\cite{QCDhamiltonian} in the weak sense (that
is, not of the
Hamiltonian itself, but a restriction thereof to the Hilbert space of
highly excited resonances, where $<k>$ is large), we note that the leading term
in the $m(k)/k$ expansion is chiral invariant, while the first order term may
involve nonchiral, spin-dependent potentials in the quark-quark interaction:
\begin{equation}
 \langle n | H^{QCD} | n' \rangle \simeq \langle n |
H^{QCD}_\chi | n'\rangle + \langle n | \frac{m(k)}{k} H^{QCD\ '}_\chi |
n' \rangle + \dots \, . \label{QCDexp}
\end{equation}

\subsection{Chiral charge and three-quark states}\label{sec:charge}

If the quarks were massless, and there would be no chiral noninvariant mass 
term in the strong interactions, the chiral charge~\cite{Nefediev:2008dv}
\begin{equation}
 Q^a_5 = \int d\vec{x} \psi^\dagger(\vec{x}) \gamma_5 \frac{\tau^a}{2}
\psi(\vec{x}) \;
\end{equation}
would commute with the QCD Hamiltonian. Nevertheless, chiral symmetry would
still be spontaneously broken by the ground state, $Q^a_5 | 0 \rangle \neq 0$,
leading to a large quark mass in the quark propagator, pseudo-Goldstone bosons
and the loss of parity-degeneracy in ground-state baryons. Using
Bogoliubov-rotated quark/antiquark operators $B$ and $D$ and the explicit
expression for the spinors, the chiral charge can be written as
\begin{eqnarray} \label{chiralcharge}
Q_5^a  = \int \frac{d^3k}{(2\pi)^3} \sum_{\lambda
\lambda ' f f'c} \left(  \frac{\tau^a}{2} \right)_{ff'}
{ k \over \sqrt{ k^2 + m^2(k)}}
 \\ \nonumber 
\times \left[ ({\bf \sigma}\cdot{\bf
\hat{k}})_{\lambda \lambda'}  	
\left( B^\dagger_{k \lambda f c} B_{k \lambda' f' c} + D^\dagger_{-k \lambda' f'
c}
D_{-k \lambda f c}
\right) + \right. \\  \nonumber \left.
{ m(k) \over k} (i\sigma_2)_{\lambda \lambda'} \
\left( B^{\dagger}_{k\lambda f c} D^\dagger_{-k\lambda'f'c}+
B_{k \lambda' f' c} D_{-k \lambda f c}
\right) \right] \ .
\end{eqnarray}
The first term between the square brackets represents a quark and antiquark
number operators flipping spin and parity. When $m(k) << k$, it dominates the
second term representing the creation or annihilation of a pion (and theRefore
realizes chiral symmetry nonlinearly).

As is argumented in Refs.~\cite{bicudoletter,Detar:1988kn,Cohen:2001gb},
repeated action of the chiral
charge on a three-quark state leads to a quartet of states, two of each parity,
which dynamically breaks into two doublets of parity partners. These partners
become degenerate when $m(k)$ vanishes. Moreover, the mass splitting between
partners is a direct measure of $m(k)$.

\section{The running quark mass and the $\Delta$
spectrum}\label{sec:runningmass}

\begin{figure}[!h]
\centering
\includegraphics[width=0.75\columnwidth]{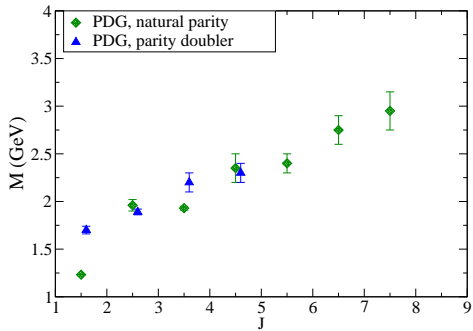}
\caption{Experimentally known masses for the lowest-lying $\Delta$ resonance for
spins up to $15/2$ and each parity. The degeneracy seen at $j=9/2$ can be
accidental and may be confirmed if the masses of the parity doublers at $j>9/2$
are measured.}
\label{fig:lowstring2}
\end{figure}

In order to link the mass splitting $|M^{P=+} - M^{P=-}|$ in a parity doublet to
the running quark mass, we will look at the lowest-lying $\Delta$ parity
doublets for increasing spin $j$ and we use the following four arguments
\begin{enumerate}
 \item Regge trajectories: $j = \alpha_0 + \alpha {M^{\pm}}^2
    \, \stackrel{_{j\to\infty}}{\longrightarrow} \, \alpha {M^{\pm}}^2$
 \item Relativistic virial theorem~\cite{virial}: $<k> \to c_2 M^\pm
\to \frac{c_2}{\sqrt{\alpha}} \sqrt{j}$
 \item The chirally invariant term ($<n|H_\chi^{QCD}|n>$) cancels out in $\Delta
M$:
 \begin{displaymath}
  |M^+-M^-| << M^\pm
 \end{displaymath}
 and
 \begin{displaymath}
      |M^+-M^-| \to \,
      <\frac{m(k)}{k} {H_\chi^{QCD}}'> \, \to \, c_3 \frac{m(k)}{k}
      <{H_\chi^{QCD}}'>
 \end{displaymath}
 \item In $H_\chi^{QCD}$, the spin-orbit $\vec{L}_i \cdot \vec{S}_i$
    term is crucial to correct the angular momentum in the centrifugal
    barrier term from $\vec{L}_i^2$ to the chirally invariant
    $\vec{L}_i^2 + 2\vec{L}_i \cdot \vec{S}_i = \vec{J}_i^2 -
    \frac{3}{4}$. Due to the sign difference in the
    helicity-dependent term $\sim - \vec{\sigma} \cdot
      \hat{k}$ in the spinor, the spin-orbit term in
    ${H^{QCD}_\chi}'$ adds to the mass difference $\Delta M$, instead
    of cancelling out as it does for $H^{QCD}_\chi$. Since the centrifugal
barrier scales like $M^\pm$ for high $j$, the
    spin-orbit term scales with one power of $j$ less:
    \begin{equation}
     <{H^{QCD}_\chi}'> \to c_5 M^\pm j^{-1} \to
    \frac{c_5}{\sqrt{\alpha}} \sqrt{\frac{1}{j}} \, . \label{eq:tensor}
\end{equation}
\end{enumerate}
Combining these four arguments, we obtain 
\begin{equation} 
 |M^+-M^-| \to \frac{c_3 c_5}{c_2
    \sqrt{\alpha}} m(<k>) j^{-1} \; .
\end{equation}
An experimental extraction can be done by fitting the exponent $-i$ of $j$ in
the splitting 
\begin{equation} \label{key1}
|M^+ - M^-| \propto j^{-i} \ .
\end{equation}
 Then, the power-law behaviour of the
running quark mass is given by 
\begin{equation} \label{key2}
m(k) \propto k^{-2i+2}\ . 
\end{equation}
The experimentally
known masses of lowest-lying $\Delta$ resonances for $j=1/2,\cdots,15/2$ are
shown in Fig.~\ref{fig:lowstring2}. From this, it is clear that the present
state of our experimental knowledge of the $\Delta$-spectrum is not
sufficient to derive the exponent $i$. Knowledge of the masses of the parity
doubler for spins $j>9/2$ would greatly enhance this.

\section{A basis for $\Delta$ states}\label{sec:basis}

The idea of extracting the running quark
mass from the hadron spectrum can be investigated using a chirally invariant
quark model. The Hamiltonian we have used in Ref.~\cite{bicudoletter} comes
from a field theory upgrade of the Cornell model and we use a linear
quark-quark potential $V_L(r) = \sigma r$ with string
tension $\sigma = 0.135$~GeV$^2$. This leads to an analytic expression for the
mass splitting between parity partners of 
\begin{eqnarray} \! \!  \! \!
M_+-M_- = 3 \int \frac{d^3k_1}{(2\pi)^3} \frac{d^3k_2}{(2\pi)^3}
\left(\frac{2}{3}\right)\int \frac{d^3q}{(2\pi)^3} \\ \nonumber
\times \hat{V}(q) \frac{1}{2} \left( \frac{m(| {\bf k}_1|)}{| {\bf
    k}_1|} + \frac{m(| {\bf k}_1+ {\bf q}|)}{| {\bf k}_1+ {\bf
    q}|} \right) F^{*\lambda_1\lambda_2\lambda_3} ({\bf k}_1, {\bf
  k}_2) \\ \nonumber
\times \left( 
\mathbb{I} - 
{\bf \sigma}\hat{\bf k}_1 
 {\bf \sigma}
\widehat{{\bf k}_1+{\bf q}}
\right)_{\lambda_1 \mu_1} F^{\mu_1\lambda_2\lambda_3}
({\bf k}_1+{\bf q}, {\bf k}_2-{\bf q}  ) \, . \label{eq:massdiff}
\end{eqnarray}
Here $\hat{V}$ is the Fourier transform of the linear potential which
falls like $q^{-4}$, and the $F$ is the wavefunction of one of the parity
partners. This wavefunction can be obtained variationally
as done in Ref.~\cite{bicudoletter}. However, it is only possible to do this
for the lowest state for each spin and parity. Higher-lying states can be
computed by diagonalizing the Hamiltonian using a set of basis functions for
three-quark states.

Constructing an orthonormal basis of three-quark states is not an easy task.
The basis functions need to form an antisymmetrized and orthonormal set
of states of definite spin and parity. We have built this basis starting from
basis states which are a combination of harmonic-oscillator (HO) states with
parameter $\alpha$ and a $2\times 2 \times 2 = 8$-component spinor
\begin{equation}
\mathcal{B}^{\alpha}_{N,l,m_\rho,m_\lambda,R_i} (\vpr,\vpl,S_i) =
\varphi^\alpha_{n^{N,l}_\rho,l^{N,l}_\rho,m_\rho}(\vpr) \,
\varphi^\alpha_{n^{N,l}_\lambda,l^{N,l}_\lambda,m_\lambda}(\vpl) \,
\chi_{R_i}(S_i) \; , \label{eq:BF}
\end{equation}
where $\vpr$ and $\vpl$ are the relative Jacobi momenta, and the indices
are needed to denote the quantum numbers of the HO wavefunctions and the
spinor.

The following subsequent steps are taken in constructing the basis
\begin{itemize}
 \item A unitary transformation is applied to the basis functions of
Eq.~\ref{eq:BF}, using Clebsch-Gordan coefficients to ensure a fixed spin of
the basis function. The parity of the state is automatically fixed for the HO
basis functions: HO shells with even (odd) shell number $N$ give rise to
positive (negative) parity states.
\item The resulting set of basis functions are symmetrised (the colour part of
the wavefunction is taken to be antisymmetric and is left out in this
discussion) by simply summing over the six possible quark permutations. This
gives rise to an overdetermined set of non-orthogonal symmetrized
basis functions of fixed spin and parity.
\item An orthonormal set of basis functions is created by performing a
Gram-Schmidt procedure on the non-orthogonal set. For this, one has to compute
the overlaps between the non-orthogonal wavefunctions using the
van~Beveren-Ribeiro-Moschinsky coefficients~\cite{beveren}.
\end{itemize}

\begin{figure}[!h]
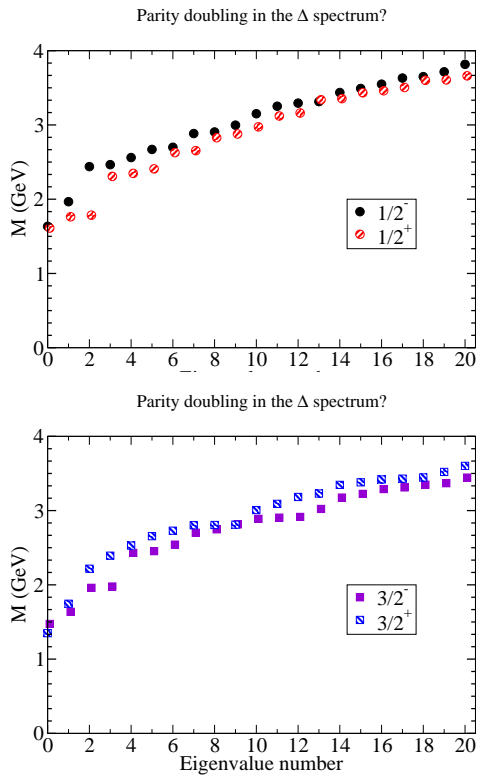

\centering
\includegraphics[width=0.75\columnwidth]{VanCauterenT-fig3.eps}
\includegraphics[width=0.75\columnwidth]{VanCauterenT-fig4.eps}
\caption{A massive variational-Montecarlo model calculation~\cite{Hahn:2004fe}
allows to check that parity doubling occurs high in the high baryon spectrum,
here $\Delta$ baryons with spin $1/2$
and $3/2$. The computation of highly excited states is currently impossible in
lattice gauge theory. Even in our simplified chiral field-theory model of QCD,
the present computation is formidable and required $\sim 10^5$ CPU-hours
at 2-3 GHz.
\label{fig:excitedstates}}
\end{figure}

We have performed this orthonormalization procedure and diagonalized the model
Hamiltonian for spin $1/2$ and $3/2$ states of each parity. The resulting
masses are plotted in Fig.~\ref{fig:excitedstates}. Here, we can already see a
tendency of
decreasing mass difference between states of different parity when looking at
higher-excited states. 

However, it is at this point not clear which states are
parity partners since the density of states is so high. In order to disentangle
the pairs in model terms, one just needs to check out the overlaps of the chiral
charge between positive and negative parity candidates. If $\langle B^-
\arrowvert Q_5^a \arrowvert B^+ \rangle\sim 1$, the states are close to being
such degenerate partners. However this is difficult to implement with
experimental data alone.

 Thus we advocate reaching high excitation not by ``radial'' excitations (successive resonances of increasing mass but equal quantum numbers), but instead by angular excitations (the ground state in each $J$-channel). It is straightforward to match the ground states in an experimental analysis, as long as good partial wave expansions are achieved. We look forward for JLAB and ELSA providing these excited $\Delta_J$ spectrum.

\section{Conclusions}\label{sec:conclusions}

It is commonly quoted in textbooks that symmetries in a quantum theory can be 
realized in Wigner mode (degenerate spectrum), Goldstone mode (vacuum
spontaneously breaks symmetry, Goldstone bosons present) or anomalously (the
quantum effective action has less symmetry than the classical action).
Chiral symmetry in QCD is widely believed to be in Goldstone mode.

In this note we have argued that actually, in QCD, chiral symmetry is realized
in both Goldstone mode (lower part of the spectrum) and Wigner mode (excited
states), and given a classical analogy.

We have further commented on the possibility of employing a new perturbative
regime in QCD, an expansion in powers of $m(k)/k$ for small quark masses. This
is useful for observables that (in Wigner mode) do not receive a contribution
from the first order term. Such observables are parity splittings in excited
states, their pion couplings, etc. These observables are then able to probe the
running quark mass in first order.

We have undertaken a major model calculation of excited baryon states, in the
simplest possible model that simultaneously implements chiral symmetry and yet
has
excited states (the first condition rules out non-relativistic quark models, the
second the Nambu-Jona-Lasinio model). Thus one needs to resort to global-color
models that are non-local, and we employ the field-theory upgrade of the Cornell
quark model. 
In this contribution we have shown results for highly excited $\Delta$ baryons
to more than 20 eigenvalues (out of reach for conceivable lattice calculations),
with a much larger variational basis in the hundred-more range of basis vectors.
Our computations, for both spin $1/2$ and $3/2$, clearly show the parity
doubling.

However, this avenue is not promising for experimental extraction, as it is then
difficult to match the partners since the density of states is large, and they
overlap as their width grows high in the spectrum. We advocate an experimental
measurement of the doubling for the ground-states of each angular momentum
channel at JLAB and ELSA.

With the data in hand, one can then use our $m(k)/k$ first-order expansion to
obtain the running of the quark mass in the infrared, a hitherto unaccessed
quantity in QCD. Other efforts are underway to access this interesting property
of a confined object from hadron structure data \footnote{Craig Roberts, private
communication} and a comparison should prove interesting.

\section*{Acknowledgments}

Work supported by grant numbers UCM-BSCH GR58/08 910309,
FPA2007-29115-E, FPA2008-00592, FIS2008-01323, CERN/FP /83582/2008,
POCI/FP /81933/2007, /81913/2007, PDCT/FP /63907/2005 and /63923/2005,
Spain-Portugal bilateral grant HP2006-0018 / E-56/07, as well as the
Flanders Research Foundation (FWO). Computational resources and
services used in this work were provided by Ghent University. FLE
acknowledges useful conversations with Craig Roberts, Eric Swanson and
Christoph Hanhart during the recent Bad-Honnef meetings.


\begin{thebibliography}{99}
%
\bibitem{Glozman:1999tk}
L.~Y.~Glozman, Phys. Lett. B {\bf 475} (2000) 329.
%
\bibitem{swanson}
E.~S.~Swanson, Phys. Lett. B {\bf 582}, (2004) 167.
%
\bibitem{Wagenbrunn:2006cs}
R.~F.~Wagenbrunn and L.~Y.~Glozman, Phys. Lett. B {\bf 643}, (2006) 98;
T.~D.~Cohen and L.~Y.~Glozman, Mod. Phys. Lett. A {\bf 21}, (2006) 1939;
L.~Y.~Glozman, A.~V.~Nefediev and J.~E.~F.~Ribeiro, Phys. Rev. D {\bf 72},
(2005) 094002.
%
\bibitem{pdg}
C.~Amsler {\it et al.}  [PDG], Phys. Lett. B {\bf 667}, (2008) 1.
%
\bibitem{LeYaouanc:1984dr}
A.~Le Yaouanc {\it et al.}, Phys. Rev.  D {\bf 31} (1985) 137.
%
\bibitem{Bicudo:1998bz}
P.~Bicudo {\it et al.}, Phys. Lett. B {\bf 442}, (1998) 349.
%
\bibitem{bowman}
P.O.~Bowman \textit{et al.}, Nucl. Phys. B, Proc. Suppl. \textbf{161}, (2006),
27; M.B.~Parappilly \textit{et al.}, Phys. Rev. \textbf{D73}, (2006) 054504;
S.~Furui, Few-Body Syst. \textbf{45}, (2009) 51; \textbf{46}, (2009) 73.
%
\bibitem{QCDhamiltonian}
For a discussion on $H^{\textrm{QCD}}$, a good starting point is N. H. Christ
and T. D. Lee, Phys. Rev. D \textbf{22}, (1980) 939.
%
\bibitem{Nefediev:2008dv}
A.~V.~Nefediev, J.~E.~F.~Ribeiro and A.~P.~Szczepaniak, JETP Lett. {\bf 87},
(2008) 271.
%
\bibitem{bicudoletter}
P.~Bicudo, M.~Cardoso T.~Van~Cauteren and Felipe~J.~Llanes-Estrada, Phys. Rev.
Lett. \textbf{103}, (2009) 092003.
%
\bibitem{Detar:1988kn}
C.~E.~DeTar and T.~Kunihiro, Phys. Rev. D {\bf 39}, (1989) 2805;
D.~Jido, T.~Hatsuda and T.~Kunihiro, Phys. Rev. Lett. {\bf 84}, (2000)
3252.
%
\bibitem{Cohen:2001gb}
T.~D.~Cohen and L.~Y.~Glozman, Phys. Rev. D {\bf 65}, (2001)
016006.
%
\bibitem{virial}
W.~Lucha and F.~F.~Schoberl, Mod. Phys. Lett. A {\bf 5}, (1990) 2473.
%
\bibitem{beveren}
E.~van~Beveren, Z. Phys. C {\bf 17}, (1983) 135.
%
\bibitem{Hahn:2004fe}
T.~Hahn, Comput.\ Phys.\ Commun.\  {\bf 168}, (2005) 78.
%
\end{thebibliography}
\end{document}